\documentclass[fleqn,usenatbib]{mnras}
\usepackage{newtxtext,newtxmath}

\usepackage[T1]{fontenc}

\DeclareRobustCommand{\VAN}[3]{#2}
\let\VANthebibliography\thebibliography
\def\thebibliography{\DeclareRobustCommand{\VAN}[3]{##3}\VANthebibliography}

\usepackage{graphicx}	
\usepackage{amsmath}	

\graphicspath{{./}{figures/}}

\newcommand{\beq}{\begin{equation}}
\newcommand{\eeq}{\end{equation}}

\newcommand{\MSun}{\,\mathrm{M_{\odot}}}
\newcommand{\Mdot}{\,\mathrm{M_{\odot}yr^{-1}}}
\newcommand{\RSun}{\,\mathrm{R_{\odot}}}
\newcommand{\LSun}{\,\mathrm{L_{\odot}}}
\newcommand{\ZSun}{\,\mathrm{Z_{\odot}}}

\newcommand{\fEdd}{f_{\rm Edd}}

\newcommand{\acc}{\MSun\mathrm{yr}^{-1}}

\newcommand{\fact}{f_{\mathrm{Edd}}}

\title[Stellar winds and SMBH seed formation in NSCs]{Effect of mass loss due to stellar winds on the formation of supermassive black hole seeds in dense nuclear star clusters}

\author[A. Das et al.]{
Arpan Das,$^{1}$\thanks{E-mail: adas45@uwo.ca}
Dominik R. G. Schleicher,$^{2}$
Shantanu Basu,$^{1}$
and Tjarda C. N. Boekholt$^{4}$
\\
$^{1}$Department of Physics and Astronomy, University of Western Ontario, London, Ontario N6A 3K7, Canada\\
$^{2}$Departamento de Astronom\'ia, Facultad Ciencias F\'isicas y Matem\'aticas, Universidad de Concepci\'on\\
$^{3}$Department of Astrophysics, American Museum of Natural History, Central Park West at 79th Street, New York, NY 10024, USA\\
$^{4}$Rudolf Peierls Centre for Theoretical Physics, Clarendon Laboratory, Parks Road, Oxford, OX1 3PU, UK
}

\date{Accepted XXX. Received YYY; in original form ZZZ}

\pubyear{2020}

\begin{document}
\label{firstpage}
\pagerange{\pageref{firstpage}--\pageref{lastpage}}
\maketitle

\begin{abstract}
The observations of high redshifts quasars at $z\gtrsim 6$ have revealed that supermassive black holes (SMBHs) of mass $\sim 10^9\,\mathrm{M_{\odot}}$ were already in place within the first $\sim$ Gyr after the Big Bang. Supermassive stars (SMSs) with masses $10^{3-5}\,\mathrm{M_{\odot}}$ are potential seeds for these observed SMBHs. A possible formation channel of these SMSs is the interplay of gas accretion and runaway stellar collisions inside dense nuclear star clusters (NSCs). However, mass loss due to stellar winds could be an important limitation for the formation of the SMSs and affect the final mass. In this paper, we study the effect of mass loss driven by stellar winds on the formation and evolution of SMSs in dense NSCs using idealised N-body simulations. Considering different accretion scenarios, we have studied the effect of the mass loss rates over a wide range of metallicities $Z_\ast=[.001-1]\mathrm{Z_{\odot}}$ and Eddington factors $f_{\rm Edd}=L_\ast/L_{\mathrm{Edd}}=0.5,0.7,\,\&\, 0.9$. For a high accretion rate of $10^{-4}\,\mathrm{M_{\odot}yr^{-1}}$, SMSs with masses $\gtrsim 10^3\MSun$ could be formed even in a high metallicity environment. For a lower accretion rate of $10^{-5}\,\mathrm{M_{\odot}yr^{-1}}$, SMSs of masses $\sim 10^{3-4}\,\mathrm{M_{\odot}}$ can be formed for all adopted values of $Z_\ast$ and $f_{\rm Edd}$, except for $Z_\ast=\mathrm{Z_{\odot}}$ and $f_{\rm Edd}=0.7$ or 0.9. For Eddington accretion, SMSs of masses $\sim 10^3\,\mathrm{M_{\odot}}$ can be formed in low metallicity environments with $Z_\ast\lesssim 0.01\mathrm{Z_{\odot}}$. The most massive SMSs of masses $\sim 10^5\,\mathrm{M_{\odot}}$ can be formed for Bondi-Hoyle accretion in environments with $Z_\ast \lesssim 0.5\mathrm{Z_{\odot}}$. An intermediate regime is likely to exist where the mass loss from the winds might no longer be relevant, while the kinetic energy deposition from the wind could still inhibit the formation of a very massive object.
\end{abstract}




\section{Introduction} \label{sec:intro}
Discovery of more than two hundred supermassive black holes (SMBHs) with masses ${\rm \gtrsim\, 10^9\MSun}$ within the first $\sim$ Gyr after the Big Bang~\citep{Fan01,Will10,Mor11,Wu15,Ban18,Mat18,Wan19,She19,Mat19} have challenged our general understanding of black hole growth and formation. How these massive objects formed and grew over cosmic time is one of the biggest puzzles to solve in astrophysics~\citep{Smi19,Ina19,Lat19}. These SMBHs are created from `seed' black holes that grow via gas accretion and mergers. The `seed' black holes are categorized into two categories: (i) low mass seeds ($\lesssim 10^2\MSun$) and (ii) high mass seeds ($\sim 10^{4-6}\MSun$). These seeds were formed at $z \sim 20-30$~\citep{Ren01}, and then they rapidly grew to their final masses by gas accretion and mergers~\citep{Day19,Pac20,Day21}. Low mass seeds are formed from Pop III stellar remnants. However, it is really challenging to grow a SMBH of mass $\sim 10^9\MSun$ from a $10^2\MSun$ seed~\citep{Hai01,Hai04,Vol12,Woo19}. A potential solution could be super-Eddington accretion \citep{Vol05,Ale14,Mad14,Vol15,Pac15,Pac17,Beg17,Toy19,Tak20}. However, radiation feedback from the seed black hole itself~\citep{Mil09,Sug18,Reg19} and inefficient gas angular momentum transfer~\citep{Ina18,Sug18} could reduce the accretion flow into the black hole.

Another solution is to start the growth from a high mass seed~\citep{Bro03}. A possible scenario for the formation of these high mass seeds is the formation of massive black holes via direct collapse \citep{Oh02,Bro03,Beg06,Aga12,Lat13,Dji14,Fer14,Bas19}. A key requirement for this scenario is large inflow rate of $\sim 0.1\MSun\text{yr}^{-1}$ which can be obtained easily in metal free halos~\citep{Aga12,Lat13,Shl16,Reg18,Bec18,Cho18,Aga19,Lat20}. In this scenario supermassive stars (SMSs) of masses $\sim 10^{4-5}\,\MSun$ are formed, which are massive enough to grow to $10^9\MSun$ by $z\sim 7$. These SMSs collapse into seed BHs with minimal mass loss at the end of their lifetime~\citep{Ume16}. These seed BHs are massive enough to grow up to $\sim 10^{9-10}\MSun$ SMBHs observed at $z\sim 7$ via Eddington accretion. The SMSs are often assumed to be formed in primordial halos with virial temperatures $T_{\mathrm{vir}}\sim10^4$ K where the formation of molecular hydrogen is fully suppressed by strong external radiation from nearby galaxies~\citep{Omu01,Sha10,Reg14,Sug14}. The accretion flow into the cloud remains very high ($0.1-1\acc$) due to the high gas temperature~\citep{Lat13,Ina14,Bec15}. The surface of the protostars is substantially inflated due to the high accretion inflow and the effective temperature drops down to a few times $10^3$ K~\citep{Hos12,Hos13,Sch13,Woo17,Hae18}. The accretion flow can then continue without being significantly affected by the radiative feedback, which allows the protostars to grow into SMSs of masses $10^{4-5}\MSun$ within about $\sim 1$ Myr. In order to prevent the molecular ${\rm H_2}$ formation, a very high background UV radiation flux is required \citep{Lat15,Wol17}, which is very rare and optimistic but not impossible~\citep{Vis14,Dji14,Reg17,Reg20}. The formation of ${\rm H_2}$ may lead to fragmentation that would suppress this process, as would the presence of metals via metal line cooling~\citep{Omu08, Dop11, Lat16, Mor18, Cho20}. 

However, there are some scenarios in which the high velocity of the baryon gas can yield high accretion rates even in the presence of ${\rm H_2}$ and/or low metallicity. High velocities due to collisions of protogalaxies \citep{Ina15} or due to supersonic streaming motions of baryons relative to dark matter \citep{Hir17} can lead to the required high mass infall rates. Massive nuclear inflows in gas-rich galaxy mergers~\citep{May15} have also been invoked.
\citet{Cho20} have further shown that even in just slightly metal enriched halos ($Z < 10^{-3}\,\ZSun$), where fragmentation takes place, the central massive stars could be fed by the accreting gas and grow into SMSs. \citet{Reg20b} also have shown that SMSs could still be formed in atomic cooling haloes with higher metal enrichment ($Z > 10^{-3}\ZSun$) in the early universe due to inhomogeneous metal distribution. The high infall rate could be obtained by dynamical heating during rapid mass growth of low-mass halos in over-dense regions at high redshifts~\citep{Wis19}. There would still be an angular momentum barrier in all scenarios, though \citet{Sak16} have shown that  gravitational instability in a circumstellar disk can solve the angular momentum problem, leads to an episodic accretion scenario \citep{Vor13,Vor15} and is consistent with the maintenance of the required low surface temperature of the accreting SMS. Nevertheless, the DCBH scenario is optimistic in that it relies on low fragmentation \citep{Lat13} and lack of disruptive feedback, e.g. x-ray feedback that could reduce the final mass of the collapsed object~\citep{Ayk14}. 

Other scenarios for the formation of SMSs are based on runaway collisions of stars in dense stellar clusters~\citep{Zwa02,Zwa04,Fre07,Fre08,Gle09,Moe11,Lup14,Kat15,Sak17,Boe18,Rei18,Sch19,Gie20,Ali20,Das20,Ver21,Riz21,Ver21}. Both Pop III star clusters~\citep[e.g.][]{Boe18,Rei18,Sch19,Ali20} and nuclear star clusters~\citep[e.g.][]{Kat15,Das20,Nat21} are possible birthplaces of such SMSs. Many galaxies host massive NSCs \citep{Car97,Bok02,leigh12,leigh15,Geo16} in their centre with masses of $\sim 10^{4-8}\MSun$. In many galaxies (typically with masses $10^9\MSun < M < 10^{11}\MSun$), NSCs and SMBHs co-exist \citep{Set08,Gra09,Cor13,Geo16,Ngu19} including galaxies like our own \citep{Sco14}, as well as M31 \citep{Ben05} and M32 \citep{Ngu18}. Studies have also found correlations between both the SMBH mass and the NSC mass with the galaxy mass \citep{Fer06,Ros06,leigh12,sco13,Set20}. In lower-mass galaxies ($M\lesssim 10^{11}\MSun$) the NSC masses are proportional to the stellar mass of the spheroidal component. The most massive galaxies do not
contain NSCs and their galactic nuclei are inhabited by SMBHs. It is therefore motivating to explore the NSCs as possible birthplaces of SMSs. There are different proposed scenarios for the formation of black hole seeds of masses $\sim 10^{3-5}\MSun$ in NSCs. \citet{Sto17} have proposed that above a critical threshold stellar mass, the NSCs can serve as possible sites for the formation of intermediate mass black holes (IMBH) and/or a SMBH from stellar collisions, which could end up eventually as central BHs via runaway tidal encounters. They have shown that both tidal capture and tidal disruption will favour the growth
of the remnant stellar mass black holes in the NSCs. In their study, they have argued that the stellar mass black holes can grow into an intermediate mass black hole (IMBH) or SMBH via three stages of runaway growth processes. At an early stage, the mass growth is driven by the unbound stars leading to supra-exponential growth. Once the BH reaches a mass
$\sim 100\MSun$, the growth is driven by the feeding of bound stars. In this second stage, the growth of the black hole could be extremely rapid as well. At later times, the growth slows down once the seed IMBH/SMBH consumes the core of its host NSC. This type of runaway growth happens in dense nuclear stellar clusters which have been observed at lower redshifts \citep[e.g.][]{Geo16}. The growth of the BH through tidal
captures/disruption of stars has also been proposed \citep{Ale17,Boe18,Arc19}. Another possible pathway for the formation and growth of massive BH seeds in NSCs is via stellar collisions and gas accretion~\citep{Dev09,Dev10,Dav11,Das20,Nat21}. A seed BH of mass $\sim 10^{4-5}\MSun$ could be formed in this case and grow to a $10^9\MSun$ SMBH at $z\sim 6-7$. \citet{Kin06} and \citet{Kin08} have shown that rapid BH growth is favoured by low values of the spin. Several studies have also proposed that NSCs are likely formed by the mergers of smaller clusters and these merging clusters may already host an IMBH which could be brought to the NSC during the merging event~\citep{Ebi01,Kim04,Zwa06,Dev12,Dav20}. It is also likely that multiple IMBHs are being fed to the NSCs~\citep{Ebi01,Mas14}, which will form binary IMBHs that could merge and emit gravitational waves (GW)~\citep{Tam18,Rass20,Arc19,Wir20}. However, the 
GW recoil kick from the merging of the two IMBHs has to be less than the escape speed of the NSC in order to retain the merged IMBH within the NSC~\citep{Ama06,Gur06,Ama07,Arc19}. A recent study by~\citet{Ask20} has shown that the SMBH will be ejected from the NSC by the GW recoil kick if the mass ratio $\gtrsim 0.15$. This might explain why some massive galaxies contain an NSC but not an SMBH, e.g. M33. 

\citet{Das20} have shown that SMSs of masses $10^{3-5}\MSun$ could be formed in dense NSCs in low metallicity environments via runaway stellar collisions and gas accretion adopting different accretion scenarios. However, in high metallicity environments the mass loss due to stellar winds will play an important role in the formation and growth of the SMSs in the NSCs. \citet{Gle13} and \citet{Kaa19} have shown that these could significantly change the final mass of the SMS formed via collisions. In this paper, we explore the effect of mass loss due to stellar winds on the final mass of the SMSs produced in nuclear clusters via gas accretion and runaway collisions. We use the same idealised N-body setups as in~\citet{Das20} and include the mass loss due to stellar winds. We adopt the theoretical mass loss recipe given by~\citet{Vin00,Vin01}. This work is an extension of the model presented in~\citet{Das20}.

\section{Methodology}
\subsection{Initial conditions}
To model collisions and accretion in the nuclear clusters consisting of main sequence (MS) stars, we use the Astrophysical MUlti-purpose Software Environment (AMUSE) \citep{Por09,Por13,Pel13,Por18}.This is an N-body code with component libraries that can be downloaded for free from \url{amusecode.org}. In the AMUSE framework
different codes, e.g. stellar dynamics, stellar evolution, hydrodynamics and
radiative transfer, can be easily coupled. We have modified the code and introduced the mass-radius relation for MS stars, gravitational N-body dynamics, gravitational coupling between the stars and the gas described via an analytic potential, accretion physics, stellar collisions, and mass growth due to accretion and collisions \citep{Das20}.

The cluster is embedded in a stationary natal gas cloud. Initially both the cluster and gas follow a Plummer density distribution
\beq
\rho(r)=\frac{3M_{cl}}{4\pi b^3}\left( 1 + \frac{r^2}{b^2}\right)^{-\frac{5}{2}}
\eeq
\citep{Plu11}, where $M_{\rm cl}$ is the mass of the cluster and $b$ is the Plummer length scale (or Plummer radius) that sets the size of the cluster core. We further assume that both the gas mass ($M_{\rm g}$) and gas radius ($R_{\rm g}$) are equal to the mass ($M_{\rm cl}$) and radius ($R_{\rm cl}$) of the stellar cluster. We introduce a cut-off radius, which is equal to five times the Plummer radius, after which the density is set to zero so that
the cluster remains stable. We consider a Salpeter initial mass function (IMF) \citep{Sal55} for the stars:
\beq
\xi(m)\Delta m = \xi_0 \left( \frac{m}{\MSun}\right)^{-\alpha}\left( \frac{\Delta m}{\MSun}\right),
\eeq
where  $\alpha = 2.35$ is the power-law slope of the mass function. We considered a top heavy IMF with mass range $10\, \MSun-100\, \MSun$. The main parameters for our simulations are the cluster mass $M_{\rm cl}$, cluster radius $R_{\rm cl}$, the gas mass $M_{\rm g}$, gas radius $R_{\rm g}$, and the number of stars $N$. 

In principle, pure N-body codes solve Newton’s equations of motion with no free physical parameters. However, they have capacities to flag special events e.g. close encounters or binary dynamics. The time-stepping criterion used to integrate the equations of
motion is the only adjustable quantity.  We used ph4 \citep[e.g.][Sec. 3.2]{Mcm96}, which is based on a fourth-order Hermite algorithm~\citep{Mak92}, to model the gravitational interactions between the stars. We modeled the gravitational effect of the gas cloud via an analytical background potential which is coupled to the $N$-body code using the BRIDGE method \citep{Fuj07}. This allows us to determine the motions of the stars from the total combined potential of the gas and stars.

\subsection{Stellar properties}
Another key ingredient in our simulations is the mass-radius relation of the MS stars as the size of the stars will play an important role in determining the number of collisions via the collisional cross section. The mass-radius ($M_\ast-R_\ast$) relation of the stars is given by
\begin{eqnarray}
\label{less50}
\frac{R_\ast}{\RSun}&=&1.60\times\left(\frac{M_\ast}{\MSun} \right)^{0.47}\,\,\,\, \mathrm{for}\,\, 10\MSun\lesssim M_\ast<50 \MSun,\\
\label{great50}
\frac{R_\ast}{\RSun}&=&0.85\times\left(\frac{M_\ast}{\MSun} \right)^{0.67}\,\,\,\, \mathrm{for}\,\, 50\MSun\lesssim M_\ast,
\end{eqnarray}
where Eq. \ref{less50} is adopted from \citet{Bon84} and Eq. \ref{great50} is adopted from \citet{Dem91}. However, it is important to note that the mass-radius relation of the SMSs is poorly understood. Moreover, stars produced via collisions could have a larger radii than similar mass stars~\citep[e.g.][]{Lom03}. Using smoothed particle hydrodynamics (SPH),~\citet{Suz07} have shown that the collision product of massive stars ($\gtrsim 100\MSun$) could be $10-100$ times larger than the equilibrium radius and hence the collision rate could be sufficiently high to have the next collision before the star settles down to the equilibrium radius. 

The luminosity of the stars is given by 
\begin{eqnarray}
\label{less120}
L_\ast&=&1.03\times M_\ast^{3.42}\LSun\,\,\,\, \mathrm{for}\,\, 10\MSun\lesssim M_\ast<120 \MSun,\\
\label{great120}
L_\ast&=&\fact \times L_{\mathrm{Edd}}\,\,\,\,\,\,\,\,\,\,\,\,\,\,\,\, \mathrm{for}\,\, 120\MSun\lesssim M_\ast,
\end{eqnarray}
where
\beq
L_{\mathrm{Edd}}=3.2\times10^4\left(\frac{M_\ast}{\MSun} \right)\LSun\, .
\label{lumEdd}
\eeq
Eq. \ref{less120} is adopted from \citet{Dem91}. As we are considering stars that are accreting, one might consider that the accretion luminosity 
\beq
L_{\mathrm{acc}}=\frac{GM_\ast\dot{M}}{R_\ast}
\eeq
also contributes to the total luminosity of the stars.
\begin{figure}
    \includegraphics[width=\columnwidth]{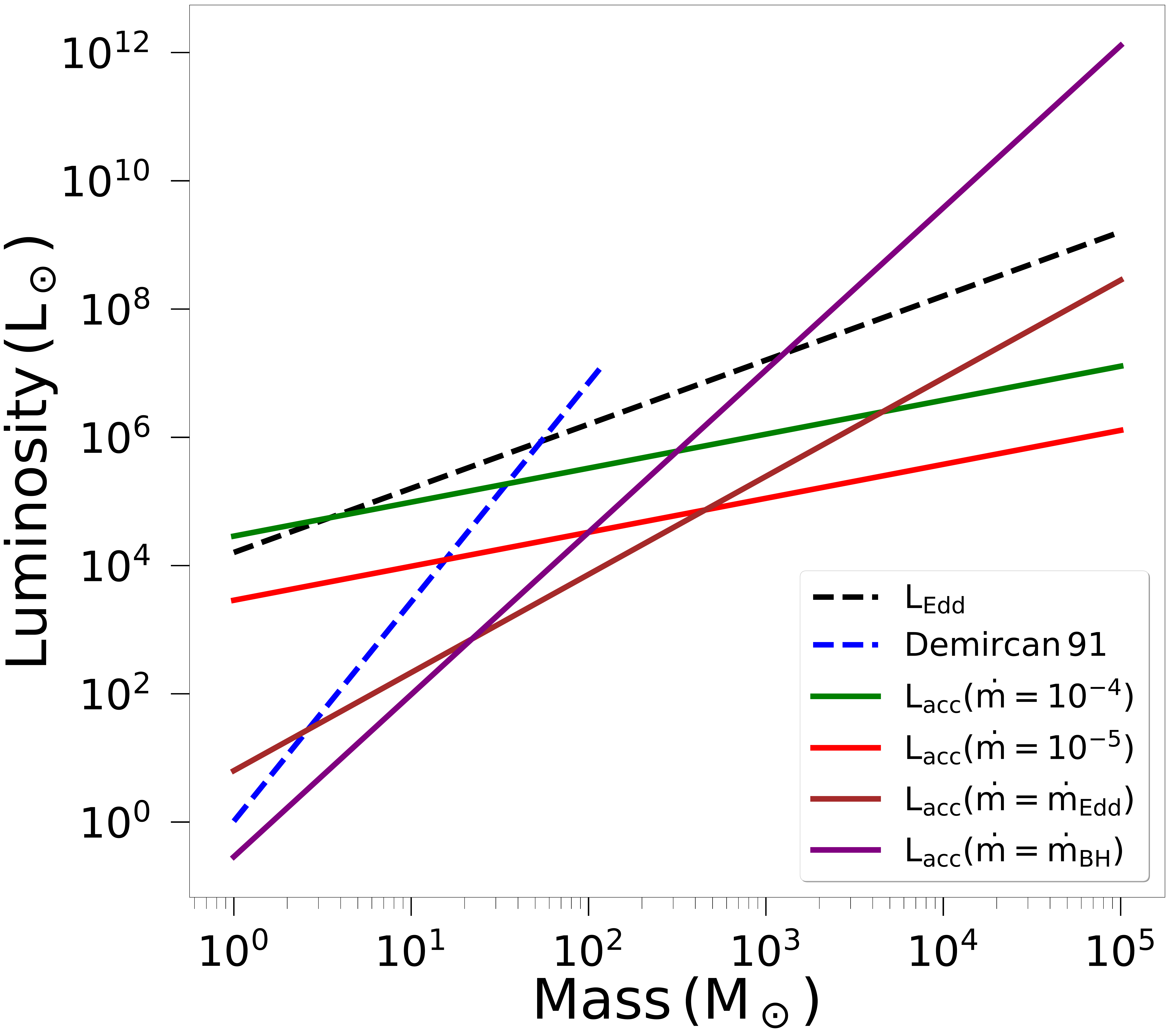}
    \caption{Luminosity of stars as a function of mass. The solid lines represent accretion luminosities whereas the dashed lines represent the Eddington luminosity and the luminosity assumed in~\citet{Dem91}, respectively.}
    \label{lum}
\end{figure}
In Fig.~\ref{lum} we have plotted the different luminosity for different accretion scenarios (see below), showing that the accretion luminosity is almost always subdominant, except for cases where we are reaching the largest stellar masses.  The atmospheric temperature of the star is given from the Stefan-Boltzman equation via
\beq
T_{\mathrm{eff}}^4=\frac{L_\ast}{4 \pi R_\ast^2 \sigma_{\mathrm{SB}}},
\eeq
where $\sigma_{SB}$ is the Stefan-Boltzmann constant.

\subsection{Gas accretion}
The next key ingredient in our simulation is the gas accretion. The protostars formed in the cluster will grow in mass by gas accretion~\citep{Bon98,Kru09,Har16}.~\citet{Das20} have found that gas accretion will play a crucial role in determining the number of collisions and hence the final mass of the most massive object (MMO).  In our current accretion prescription, the gas is assumed to be initially at rest, and hence due to momentum conservation the  stellar velocity decreases as they accrete gas and gain mass, and fall deeper into the potential well of the cluster. We assume that no new stars are formed and hence the gas is fueled into the cluster with $100\%$ accretion efficiency. The efficiency might be reduced due to protostellar outflows~\citep{Fed15,Off17}, which are not considered here. At each time step the accreted gas mass is subtracted from the total gas mass, and the density keeps being distributed according to the Plummer profile. Hence, the gas depletion in our simulation is uniform. We consider different accretion scenarios in our work, including constant accretion rates  of $10^{-4},\,10^{-5}$ and $10^{-6} \MSun\mathrm{yr^{-1}}$, and Eddington accretion given by:
\beq
\dot{M}_{\mathrm{Edd}}=2.2\times 10^{-8}\left(\frac{M_\ast}{\MSun} \right)\,\,\mathrm{\MSun\, yr^{-1}}\, ,
\label{eddacc}
\eeq
as well as Bondi-Hoyle-Lyttleton (hereafter BHL or Bondi-Hoyle) accretion given by Eq. 2 of \citet{Macc12}:
\begin{equation}
\label{mainbondi}
\dot{M}_{\mathrm{BHL}} = 7\times 10^{-9} \left(\frac{M_\ast}{\MSun}\right)^2 \left(\frac{n}{10^{6}\, \mathrm{cm}^{-3}} \right)^2\left(\frac{\sqrt{c_{\mathrm{s}}^2+v_\infty^2}}{10^6\,\mathrm{cm\ s}^{-1}} \right)^{-3}\mathrm{\MSun\, yr^{-1}}.
\end{equation}
A recent study by \citet{Kaa19} has shown that the average BH accretion rate of an individual star is given by
\begin{equation}
\label{kaaz}
  \langle \dot{M}_{\mathrm{BHL}} \rangle=\left\{
  \begin{array}{@{}ll@{}}
    \dot{M}_{\mathrm{BHL}}, & \text{when}\ R_{\bot} \gg R_{\mathrm{acc}}, \\
    N\times\dot{M}_{\mathrm{BHL}}, & \text{when}\ R_{\bot} \leq R_{\mathrm{acc}},
  \end{array}\right.
\end{equation} 
where $R_{\bot}=R_{\mathrm{cl}} N^{-1/3}$ is the mean separation between stars and $R_{\mathrm{acc}}=\frac{2GM_\ast}{v_\infty^2}$ is the characteristic accretion radius of a star, i.e. the impact parameter in the BHL theory for which gas can be gravitationally-focused and overcome its angular momentum barrier to reach the star. Here, $v_\infty$ is the relative velocity of the star with respect to the gas. Our adopted BHL accretion rate is given by Eq.~\ref{mainbondi}. We multiply the BHL rate of a single star by $N$ if $R_{\bot} \leq R_{\mathrm{acc}}$ according to Eq.~\ref{kaaz}. We compute the density of the gas and hence the $\dot{M}_{\mathrm{BHL}}$ locally.

\subsection{Stellar Collisions}
We adopt the sticky-sphere approximation to model collisions between the main sequence stars \citep{leigh12b, leigh17}, where the two stars are assumed to merge if the distance between their centres is less than the sum of their radii. The two stars are replaced by a single star whose mass is equal to the sum of the masses of the colliding stars and the radius of the object to be determined by the ($M_\ast-R_\ast$) relation described in Eqs. \ref{less50} and \ref{great50}. The conservation of linear momentum is implemented during the collision. However,  the mass is not necessarily conserved due to the possible ejection of mass~\citep{Sil02,Dal06,Tra07}. The final mass of the colliding objects could change a lot depending on fraction of the mass that is lost during the merger~\citep{Ali20,Das20}. This fraction depends on the type of stars that are colliding~\citep{Gle13}. 

\subsection{Mass loss due to stellar winds}
Since the massive stars and the collision products in our simulations become very massive and luminous, mass loss driven by stellar wind plays a key role in their evolution. However, the mass loss of very massive stars is very poorly understood both observationally and theoretically. In this work we adopt the theoretical mass loss recipe given by~\citet{Vin00,Vin01}. The mass loss rate is a function of the luminosity of the stars $L_\ast$, mass of the stars $M_\ast$, the Galactic ratio of terminal velocity and escape velocity $v_\infty/v_{\mathrm{esc}}$, the effective temperature of the stars $T_\mathrm{eff}$, and the metallicity of the stars $Z_\ast$.

\begin{figure*}
    \includegraphics[width=0.65\columnwidth]{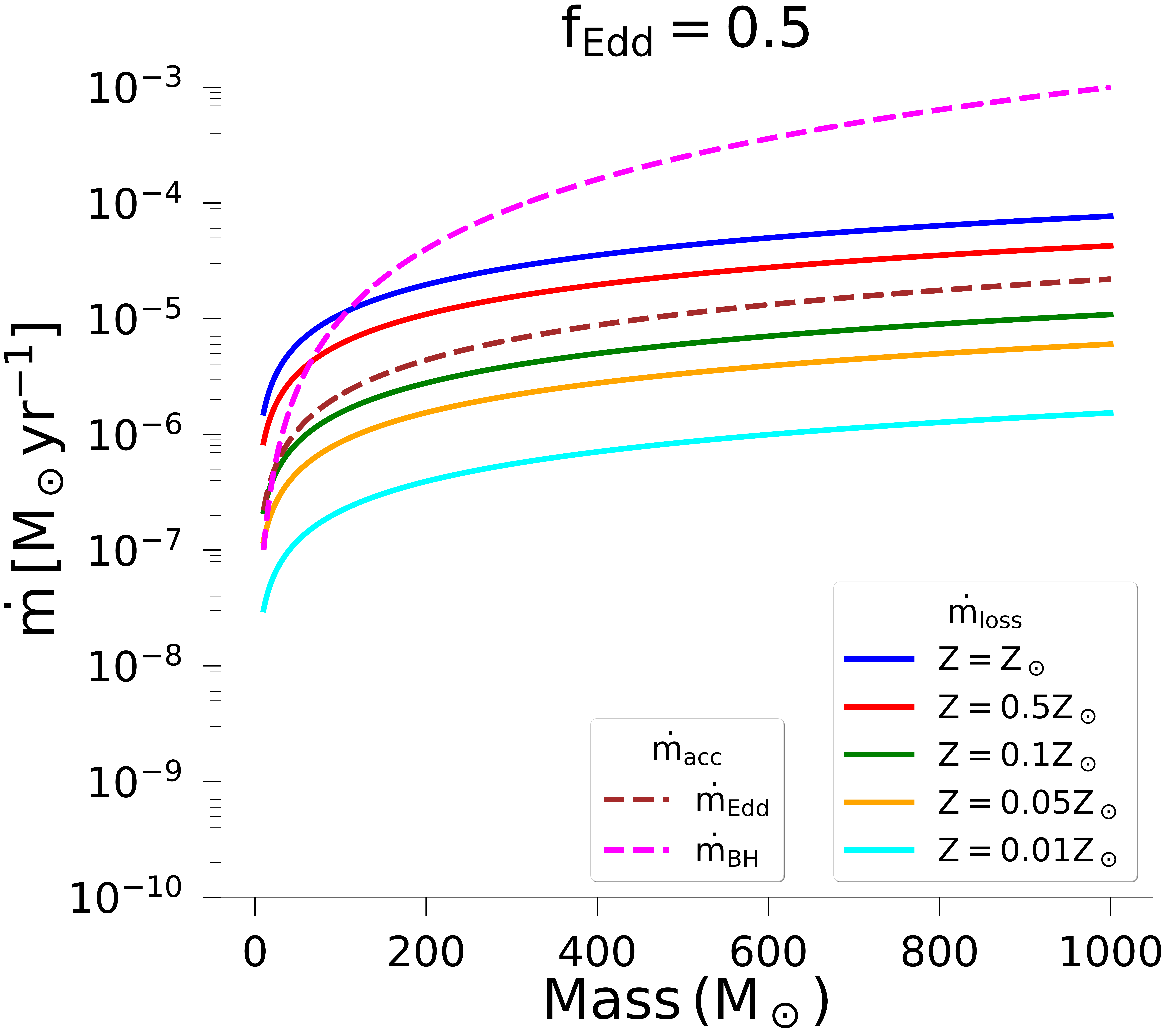}
    \includegraphics[width=0.65\columnwidth]{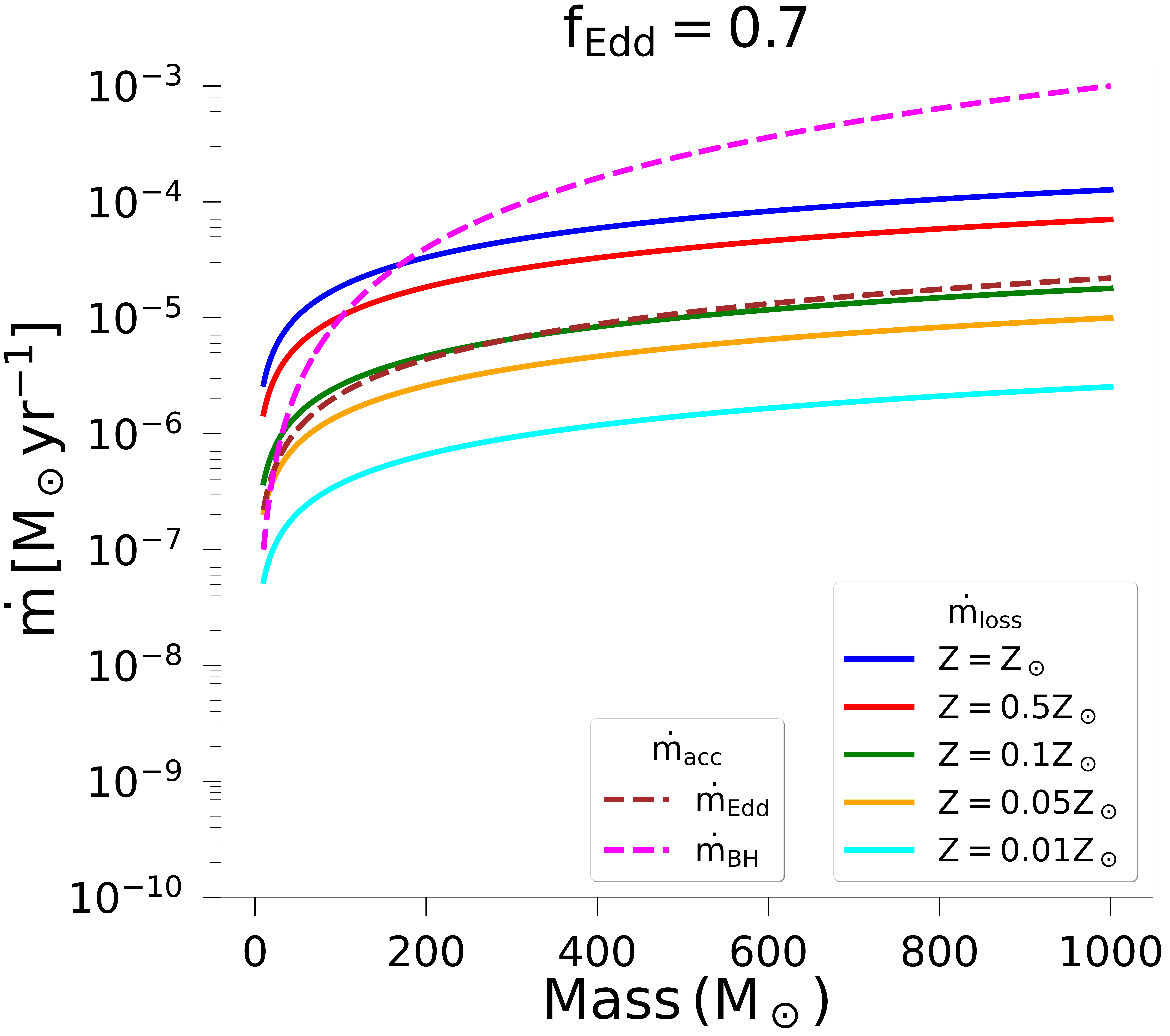}
    \includegraphics[width=0.65\columnwidth]{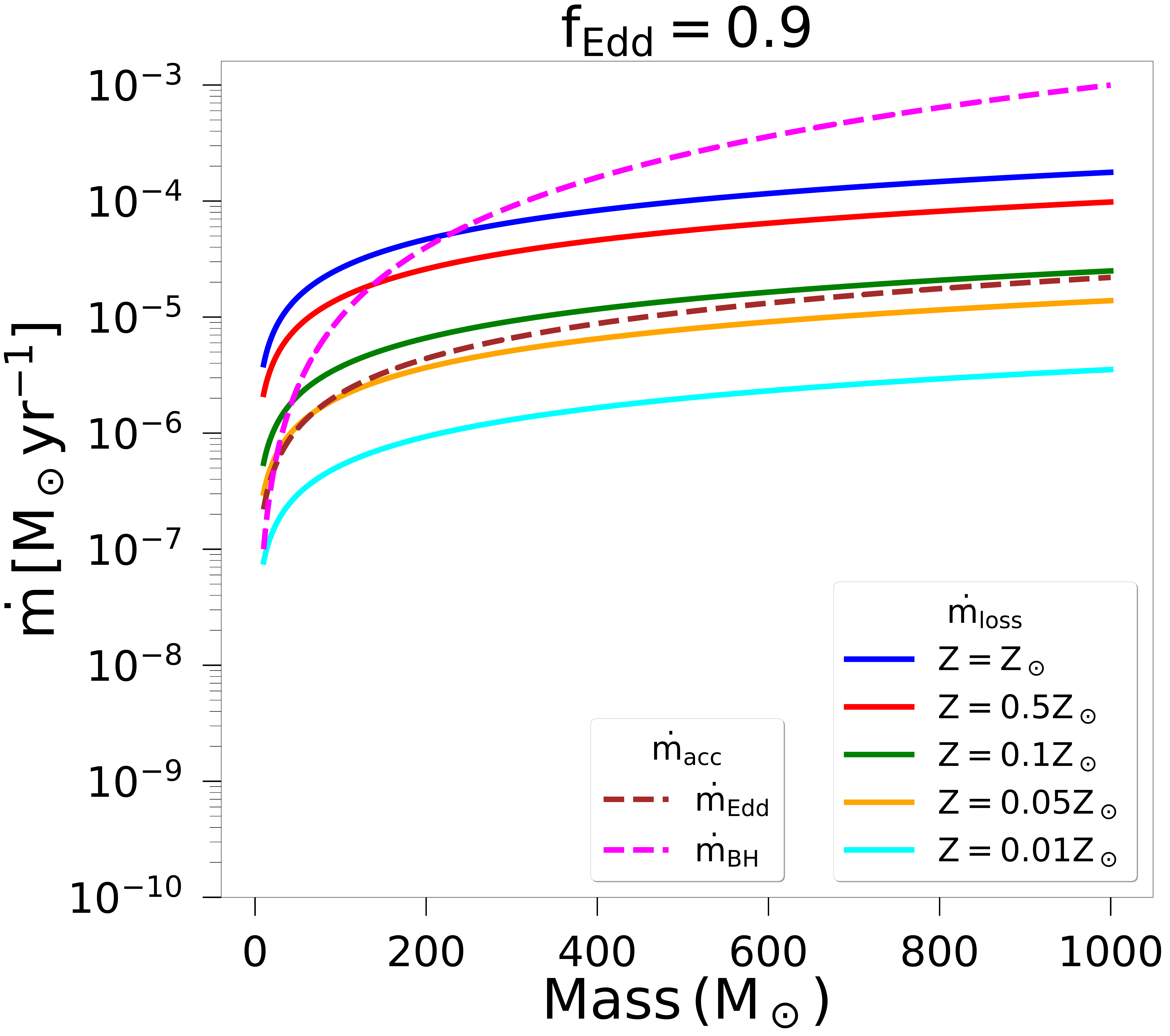}
    \caption{Mass loss rate $\dot{m}_\mathrm{loss}$ as a function of mass for different metallicities (solid lines) and accretion rates (dashed line) as a function of the stellar mass.}
\label{rates}
\end{figure*}
In Fig.~\ref{rates} we have plotted the mass loss rates for different metallicities. We have also plotted the Eddington and Bondi-Hoyle accretion rates to compare with the mass loss rates (the constant accretion rates can be identified without a plotted line). It is interesting to note that for Bondi-Hoyle accretion, the mass loss could be comparable or higher for masses $\lesssim 200\MSun$ and metallicities $Z=(0.5-1)\ZSun$, whereas the Eddington accretion rate is always lower than the mass loss rate in the same metallicity range for any mass. The Eddington accretion rate could be comparable to or higher than the mass loss rate for $Z\lesssim 0.1\ZSun$. Another key parameter in the mass loss rate is the Eddington factor $f_{\mathrm{Edd}}$, which is given by Eq.~\ref{lumEdd}. \citet{Nad05} have shown that $0.54\lesssim\fact\lesssim 0.94$ for stellar masses in the range $3\times 10^2\MSun\lesssim M_\ast\lesssim 10^4\MSun$. We adopt a typical value of $\fact=0.7$ for the rest of our models. In Fig.~\ref{rates} we show the mass loss rates for different values of $f_{\mathrm{Edd}}$. An important point to note here is that the mass loss recipe in~\citet{Vin01} was computed for $f_{\mathrm{Edd}}<0.5$. \citet{Vin11} have shown that mass loss rates could be significantly higher for stars close to the Eddington limit. In other words, when extrapolating the results to higher Eddington fractions, it is important to note that we might be underestimating the mass loss that actually occurs.

\section{Results}
The main results of our simulations are presented in this section. We adopted the initial conditions with $N=5000$, $M_{\mathrm{cl}}=M_{\mathrm{g}}=1.12\times 10^5 \MSun$, $R_{\mathrm{cl}}=R_{\mathrm{g}}=1$ pc, assuming a Salpeter IMF within a stellar mass interval of $10-100 \MSun$. We have assumed three different values of $\fEdd=0.5,0.7,0.9$ and for each $\fEdd$, we have studied six different metallicities $Z_\ast=\ZSun,\,0.5\ZSun,\,0.1\ZSun,\,0.05\ZSun,\,0.01\ZSun,\,0.001\ZSun$. 

The evolution of the cluster is similar to the results in~\citet{Das20}. In the initial phase the stars accrete gas and due to momentum conservation the stellar velocity decreases and the stars fall deeper into the potential well of the cluster. During the accretion phase the accretion dominates the mass growth. Once the gas is fully depleted, the stellar collisions take place and drive the mass growth of the SMS. However, some initial collisions might occur which will boost the accretion process, especially for 
the Eddington and Bondi-Hoyle accretion scenarios. The evolution of the Lagrangian radii will be similar to our previous results in~\citet{Das20}. The $10\%$ Lagrange radii will always decrease, eventually leading to a core collapse. The evolution of the $50\%$ and $90\%$ Lagrange radii will be an initial decrease and a later increase after the core collapse. The timing of the transition will depend on the accretion recipe. A similar trend has been seen in previous simulations in the absence of accretion \citep[e.g.][]{leigh14}.

\begin{figure*}
    \includegraphics[width=0.5\columnwidth]{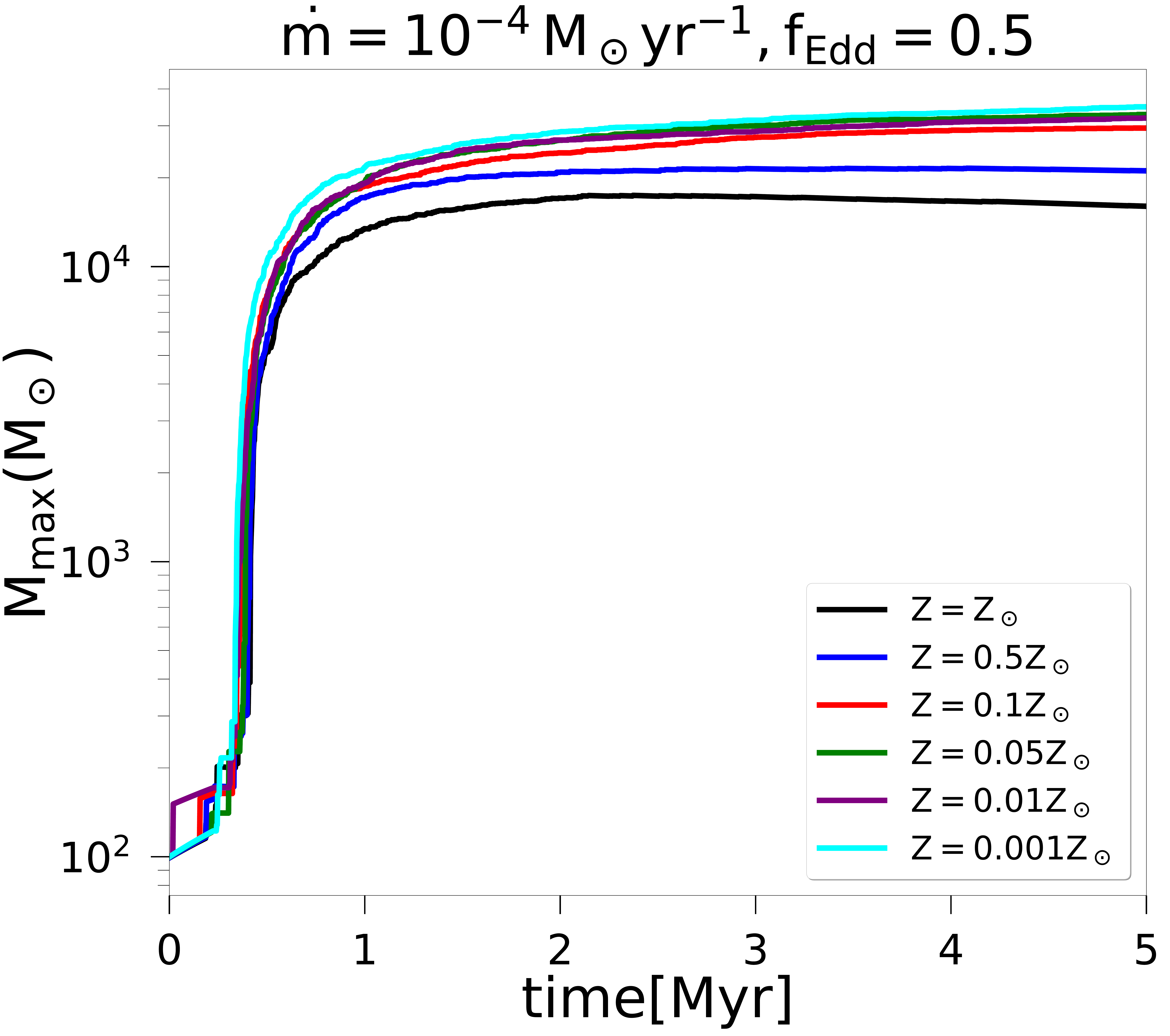}
    \includegraphics[width=0.5\columnwidth]{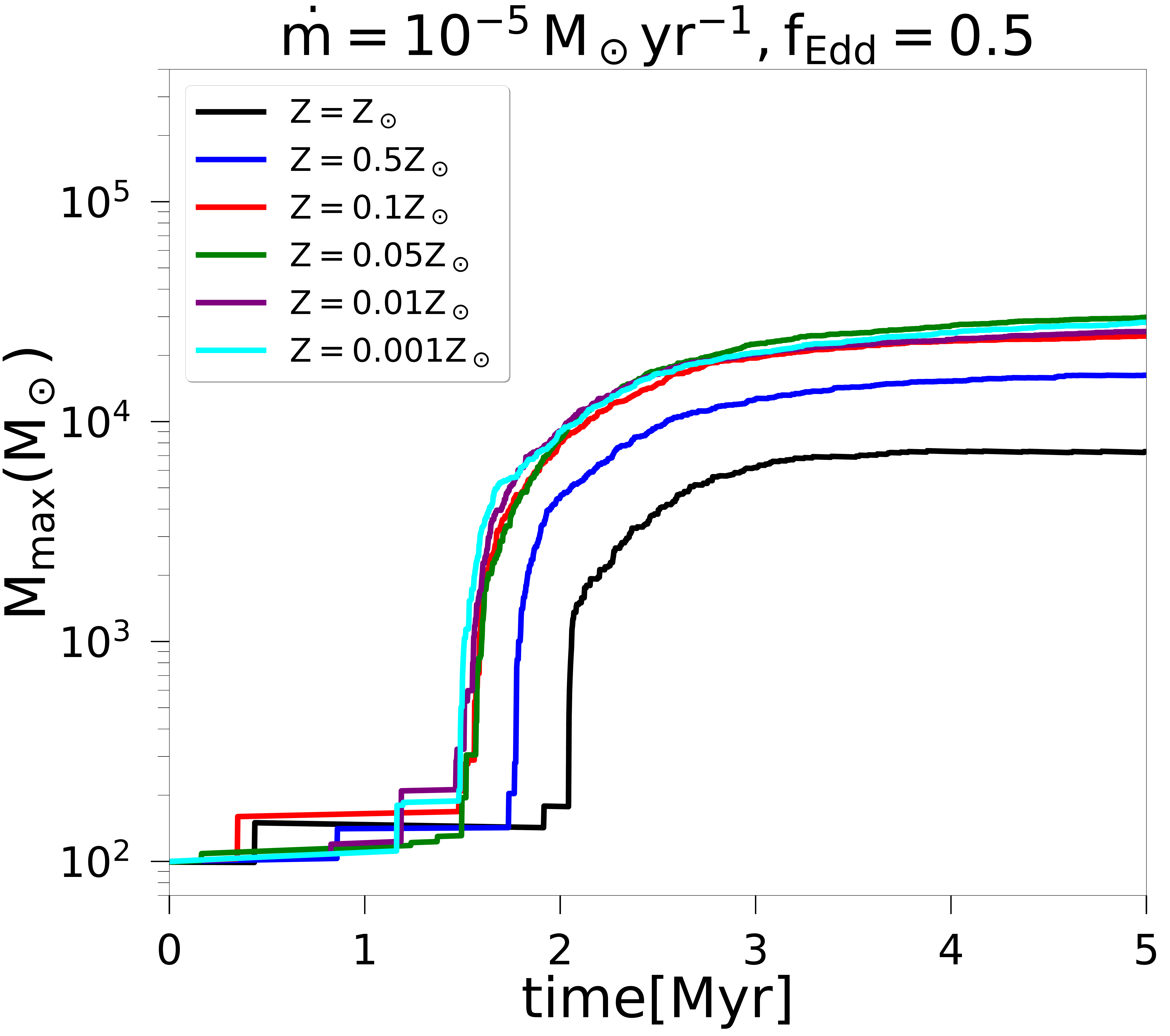}
    \includegraphics[width=0.5\columnwidth]{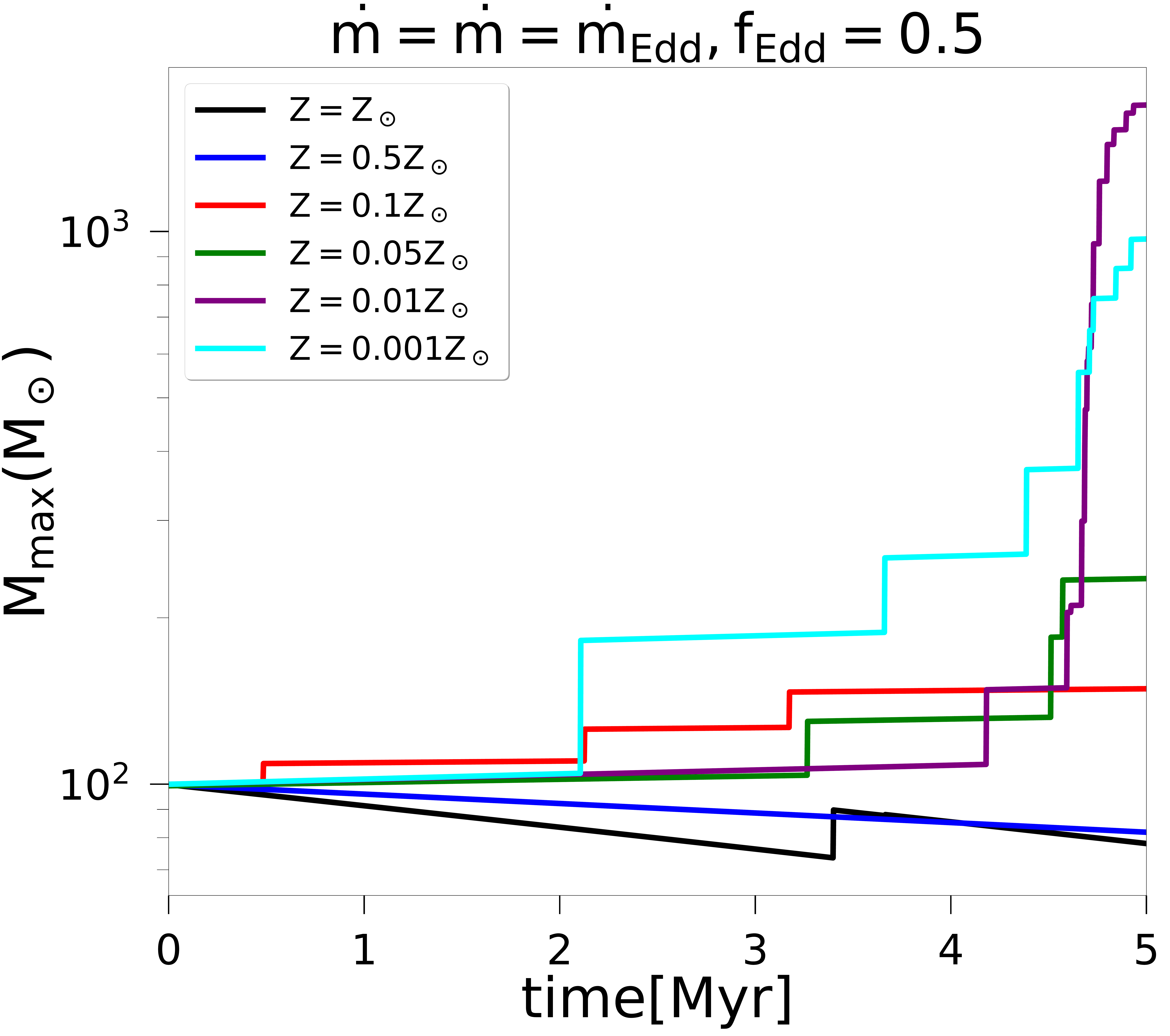}
    \includegraphics[width=0.5\columnwidth]{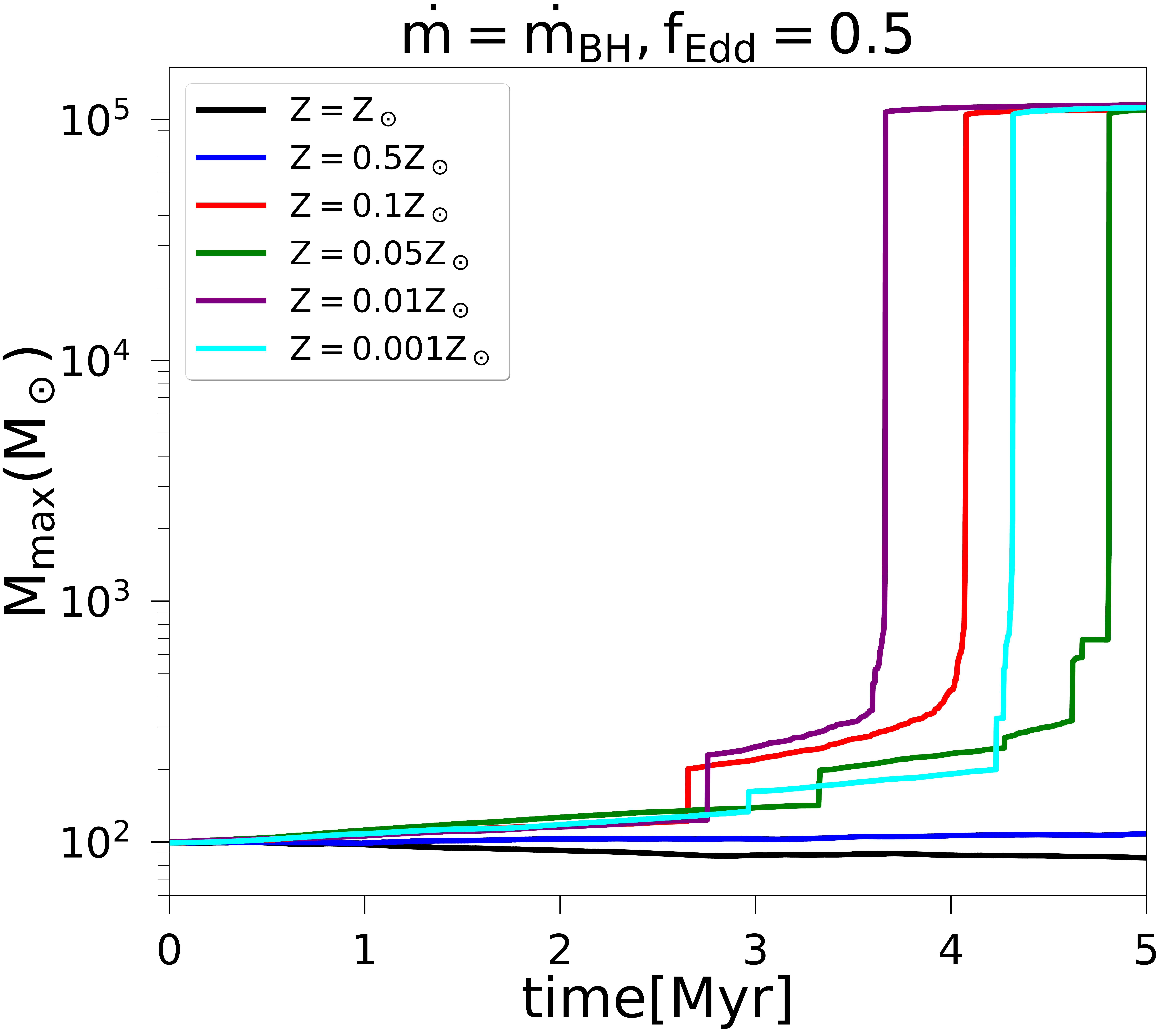}
    \includegraphics[width=0.5\columnwidth]{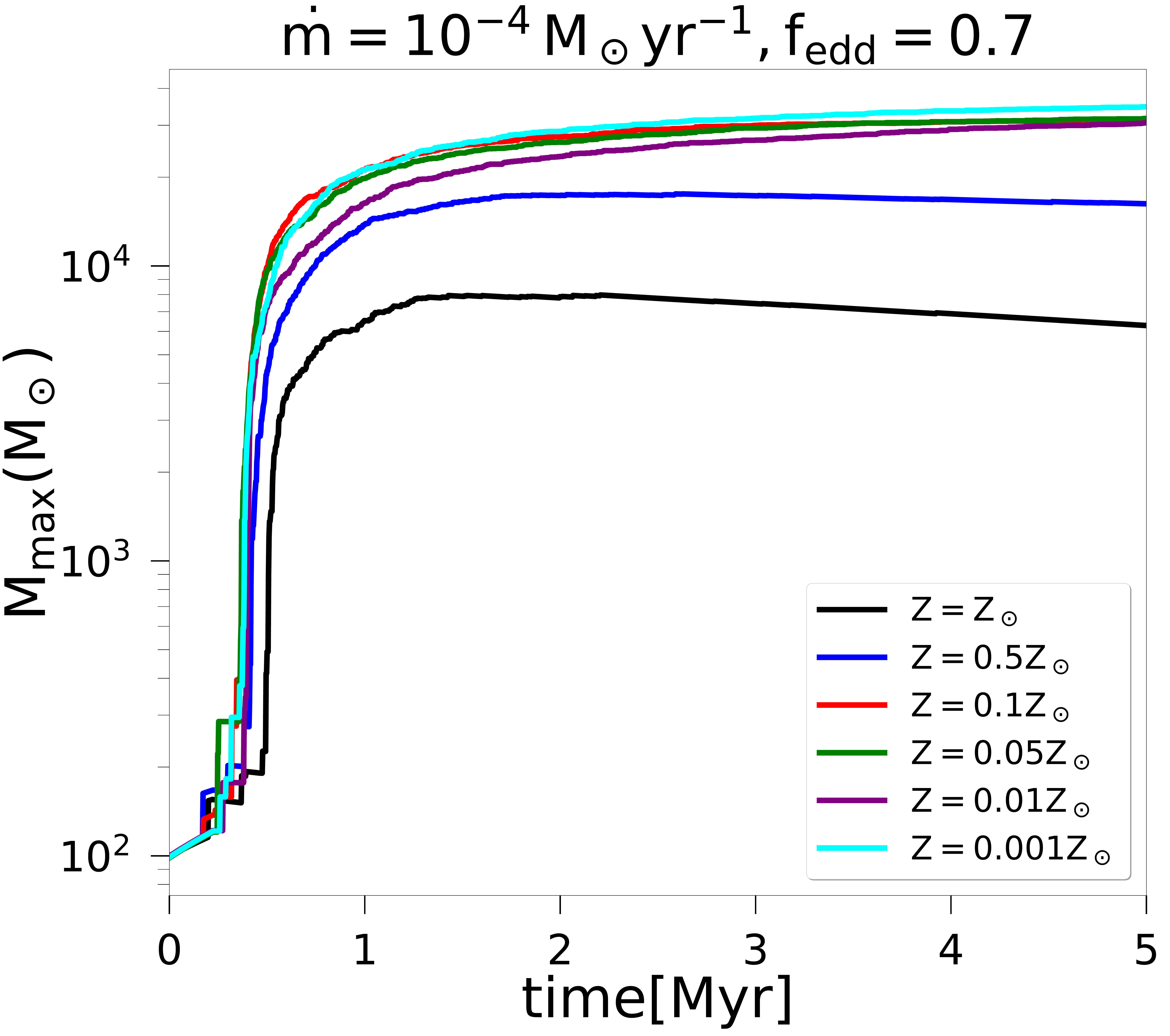}
    \includegraphics[width=0.5\columnwidth]{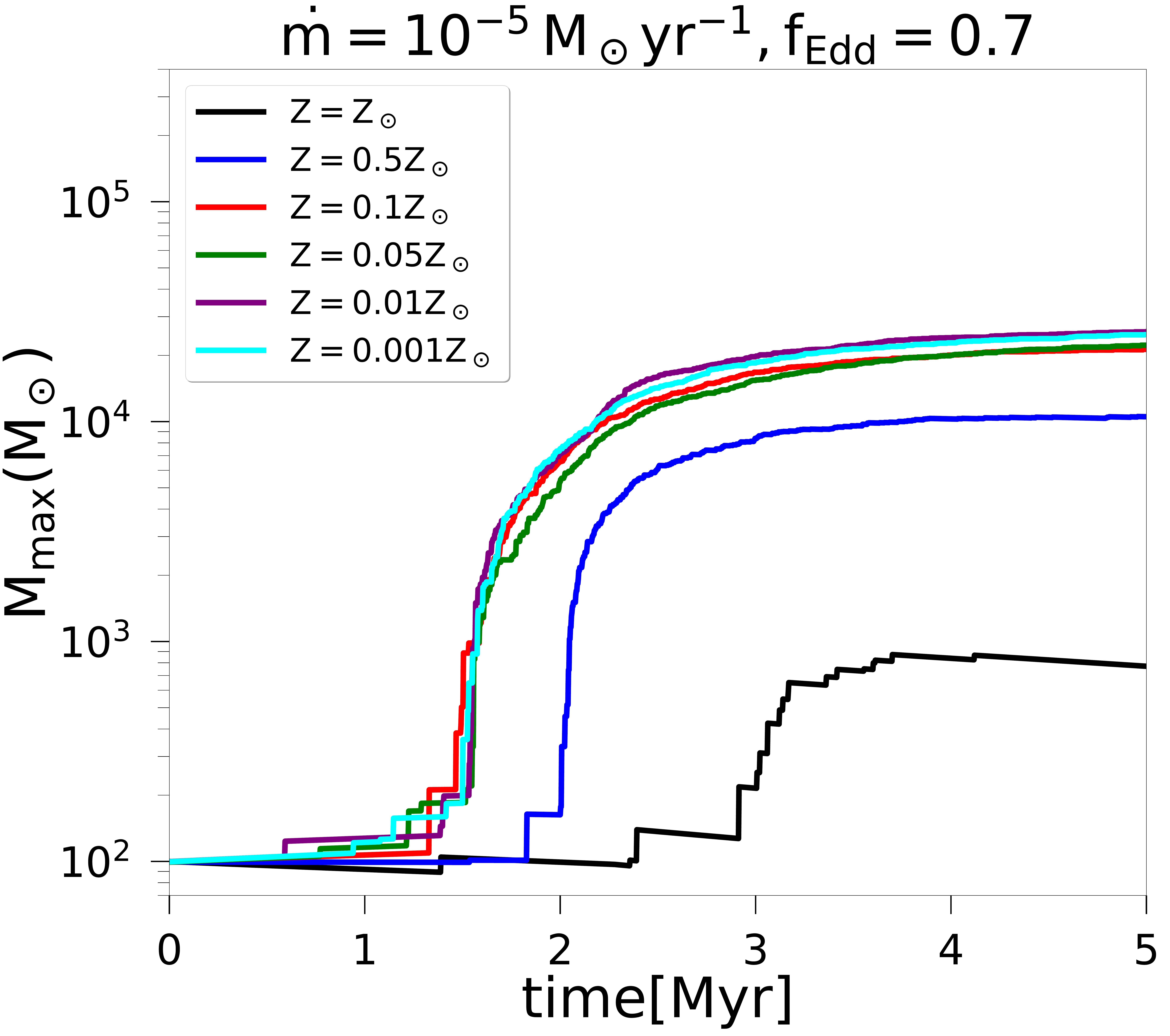}
    \includegraphics[width=0.5\columnwidth]{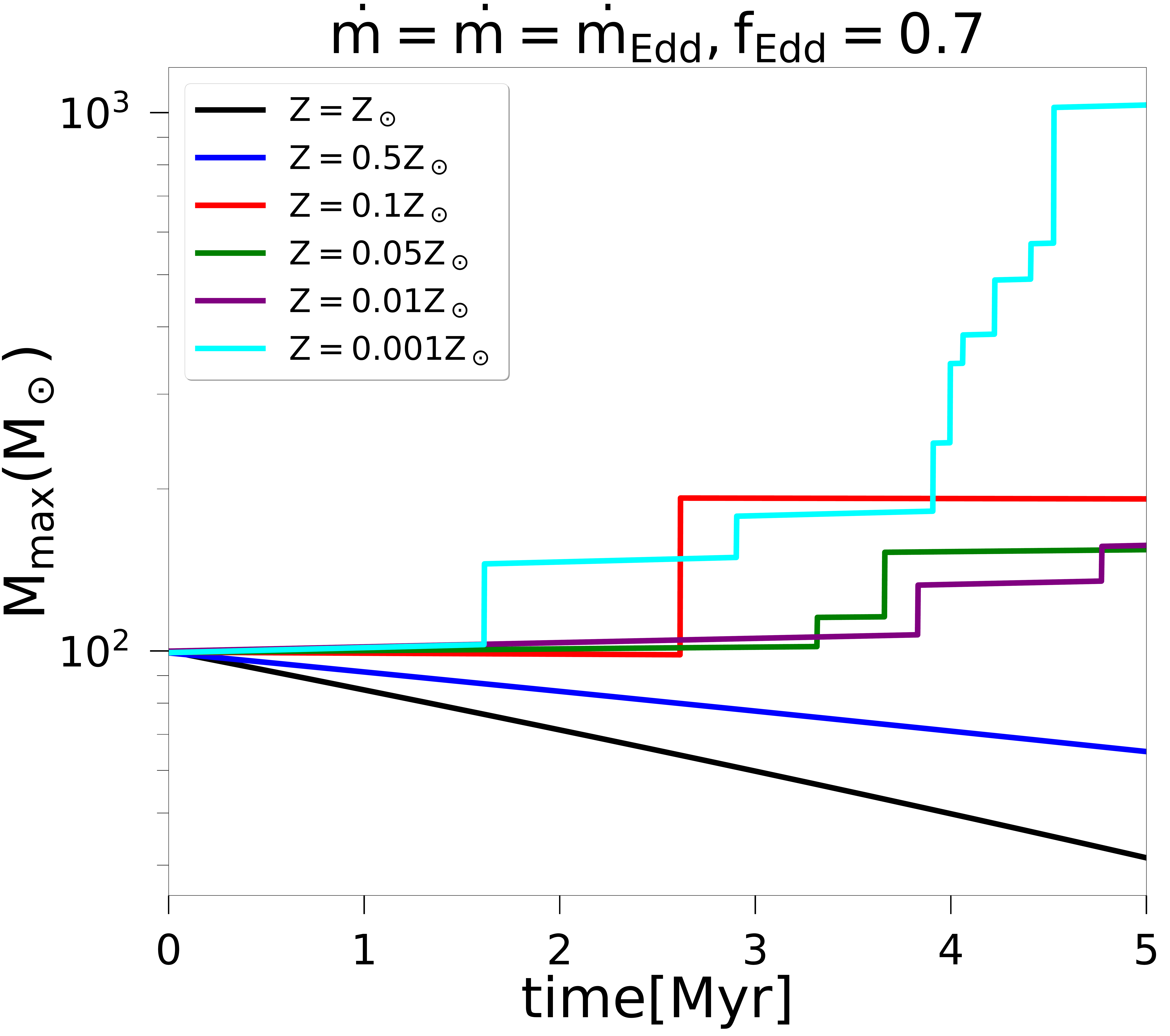}
    \includegraphics[width=0.5\columnwidth]{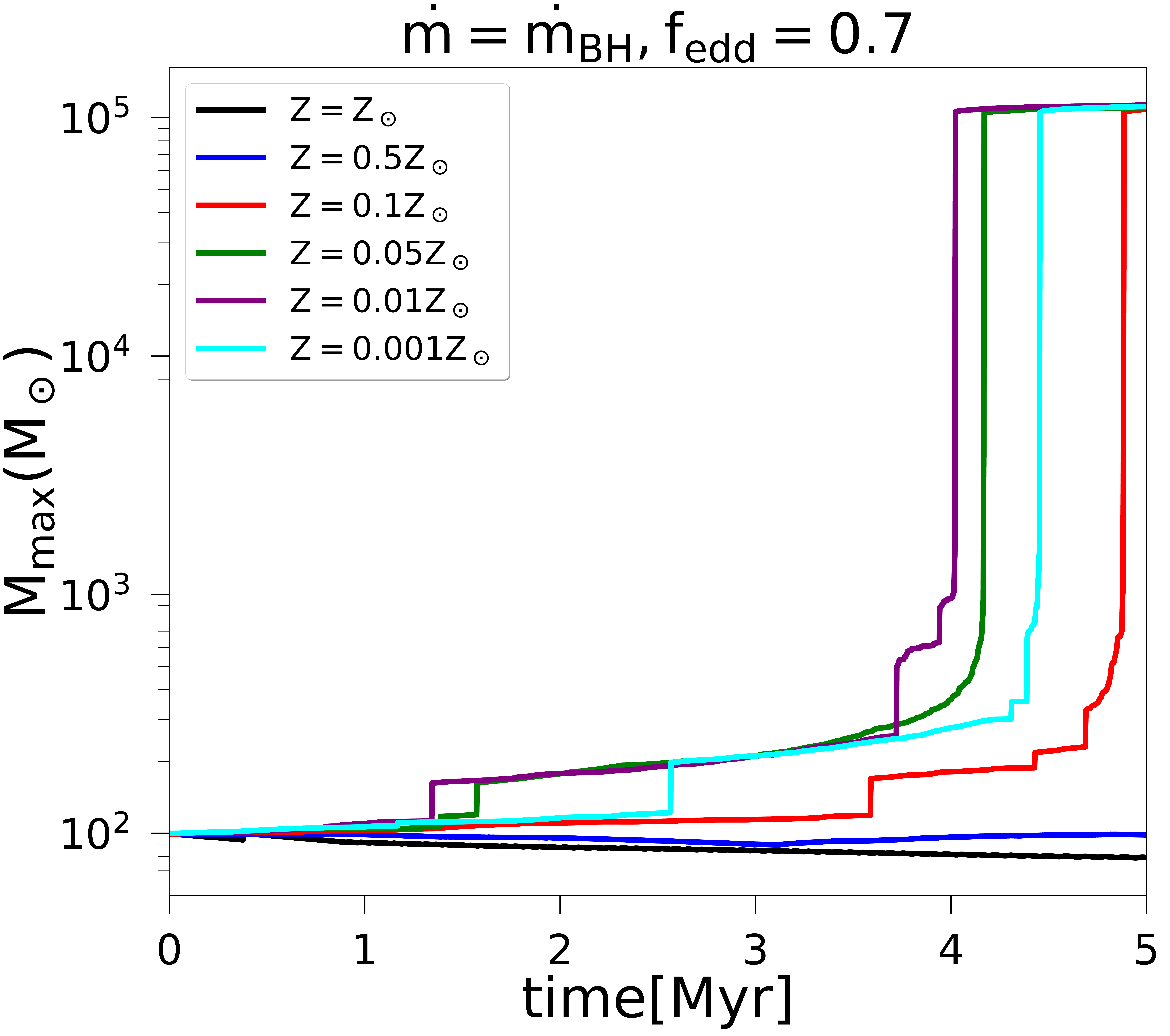}
    \includegraphics[width=0.5\columnwidth]{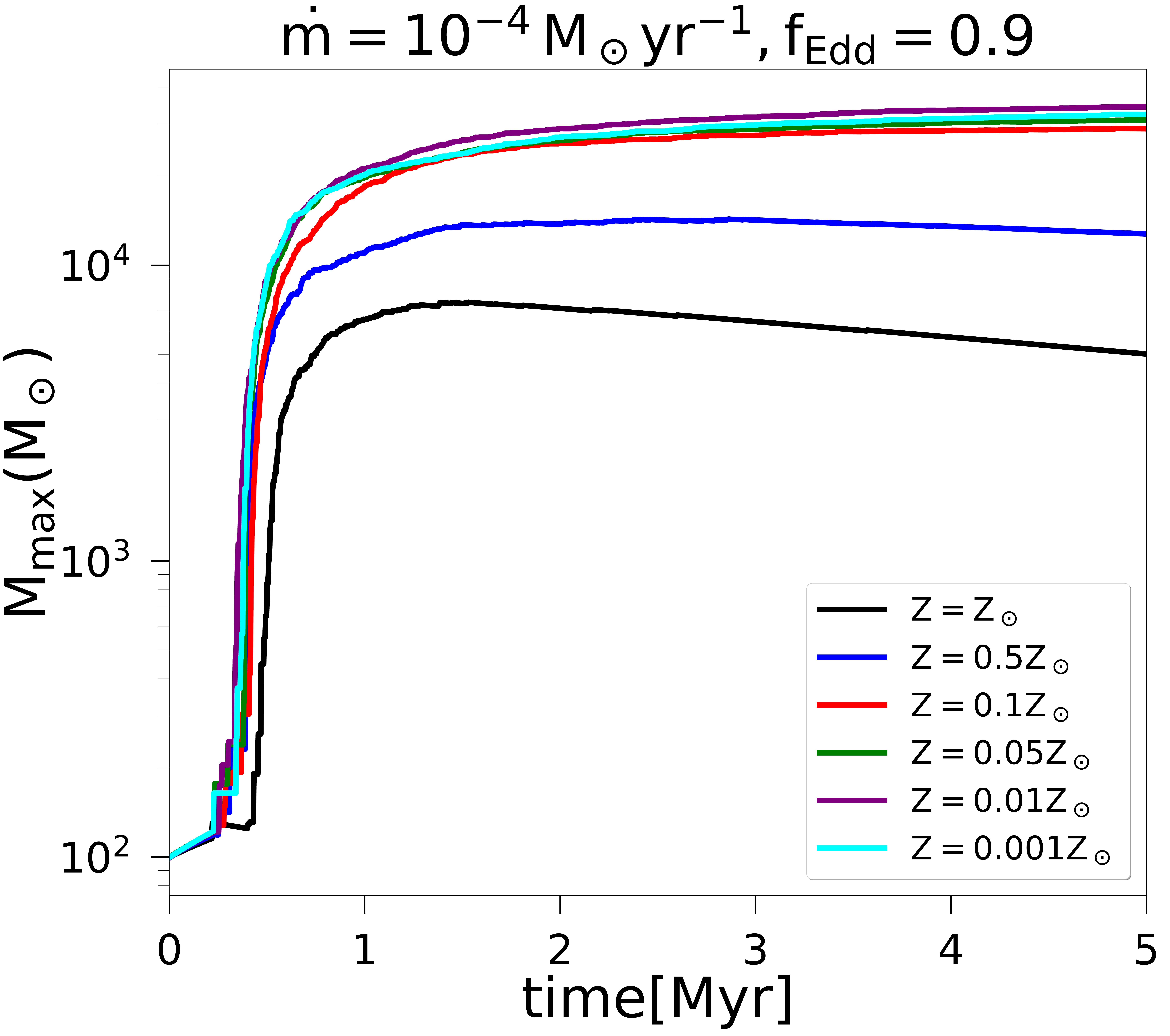}
    \includegraphics[width=0.5\columnwidth]{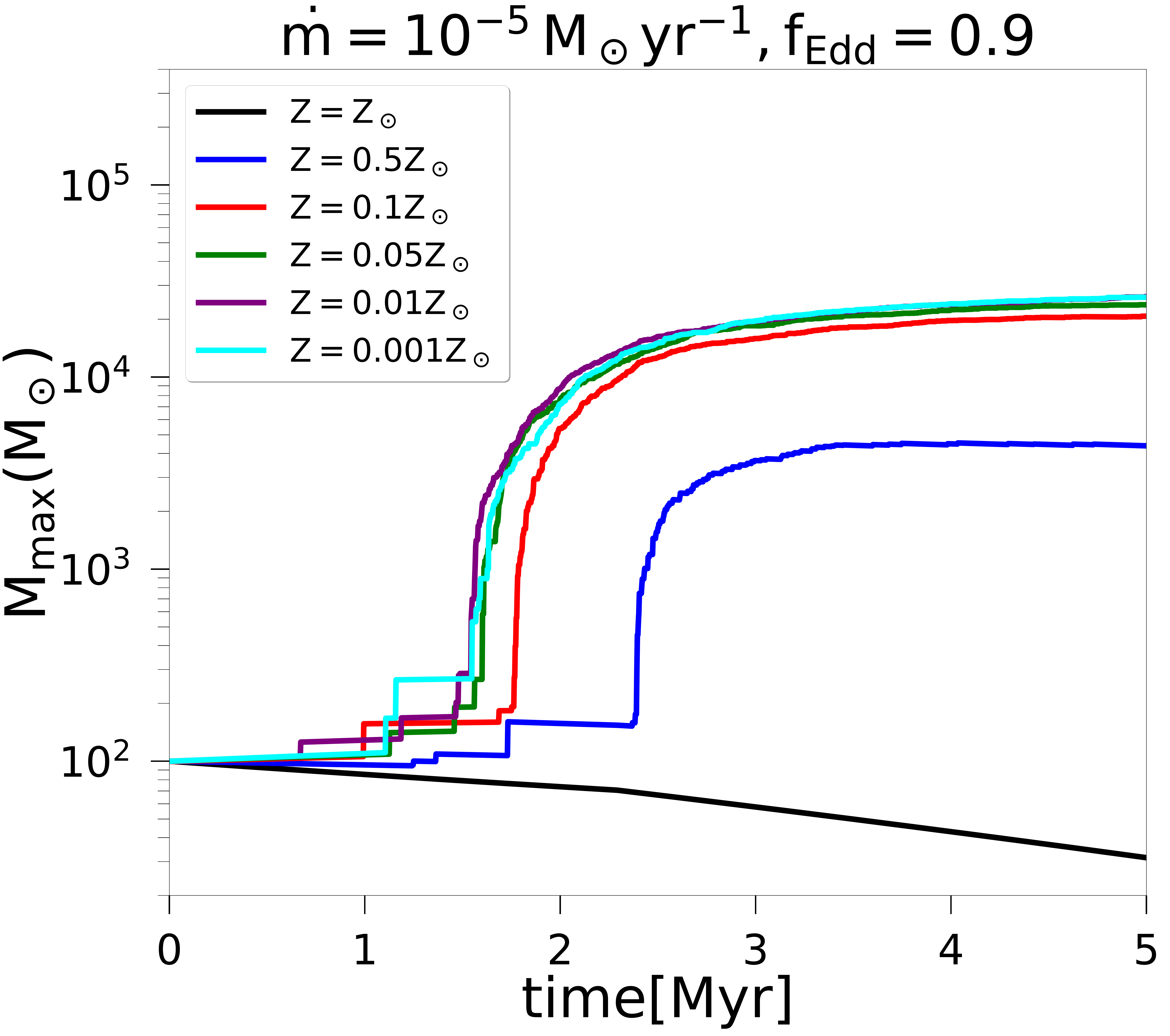}
    \includegraphics[width=0.5\columnwidth]{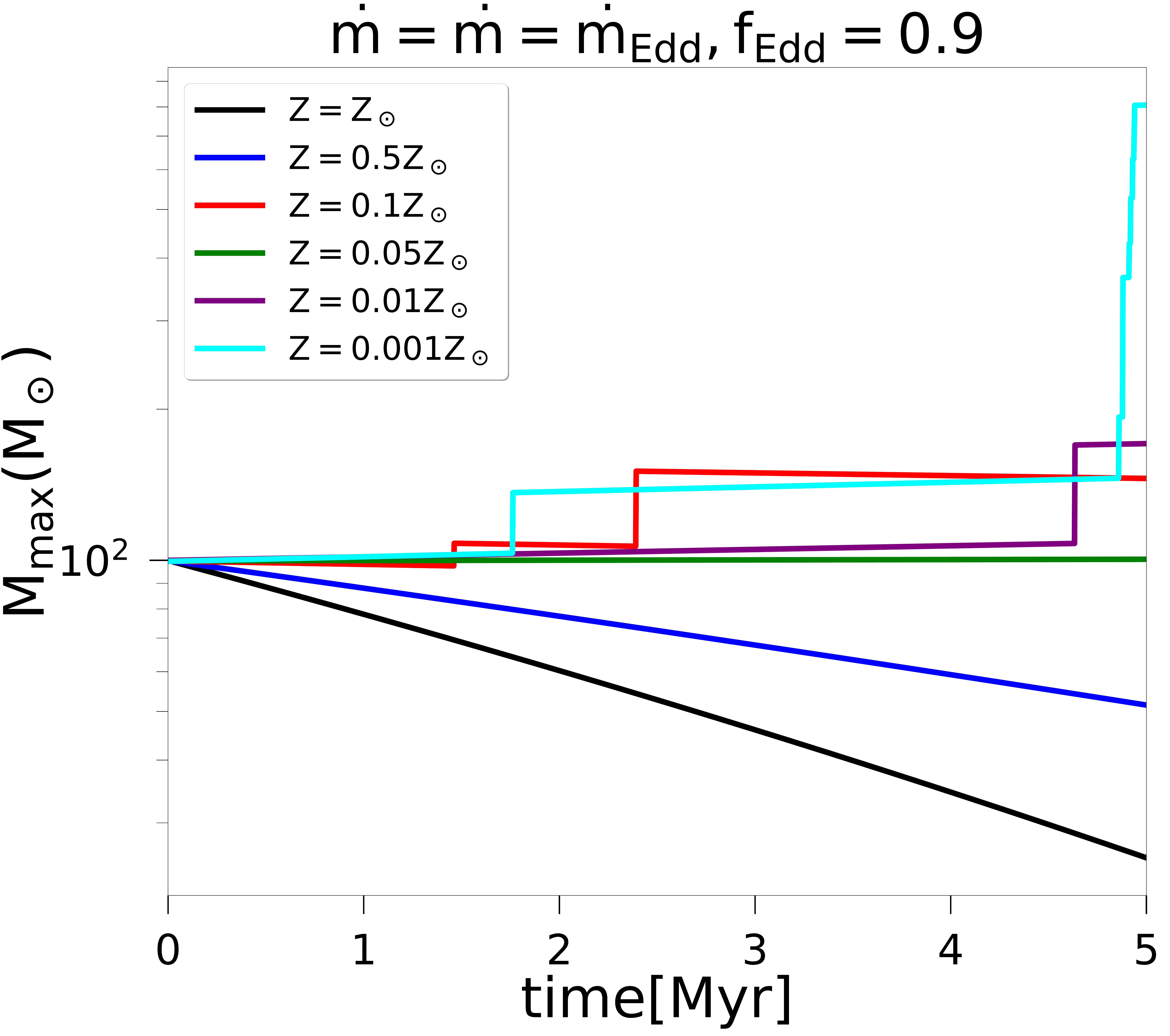}
    \includegraphics[width=0.5\columnwidth]{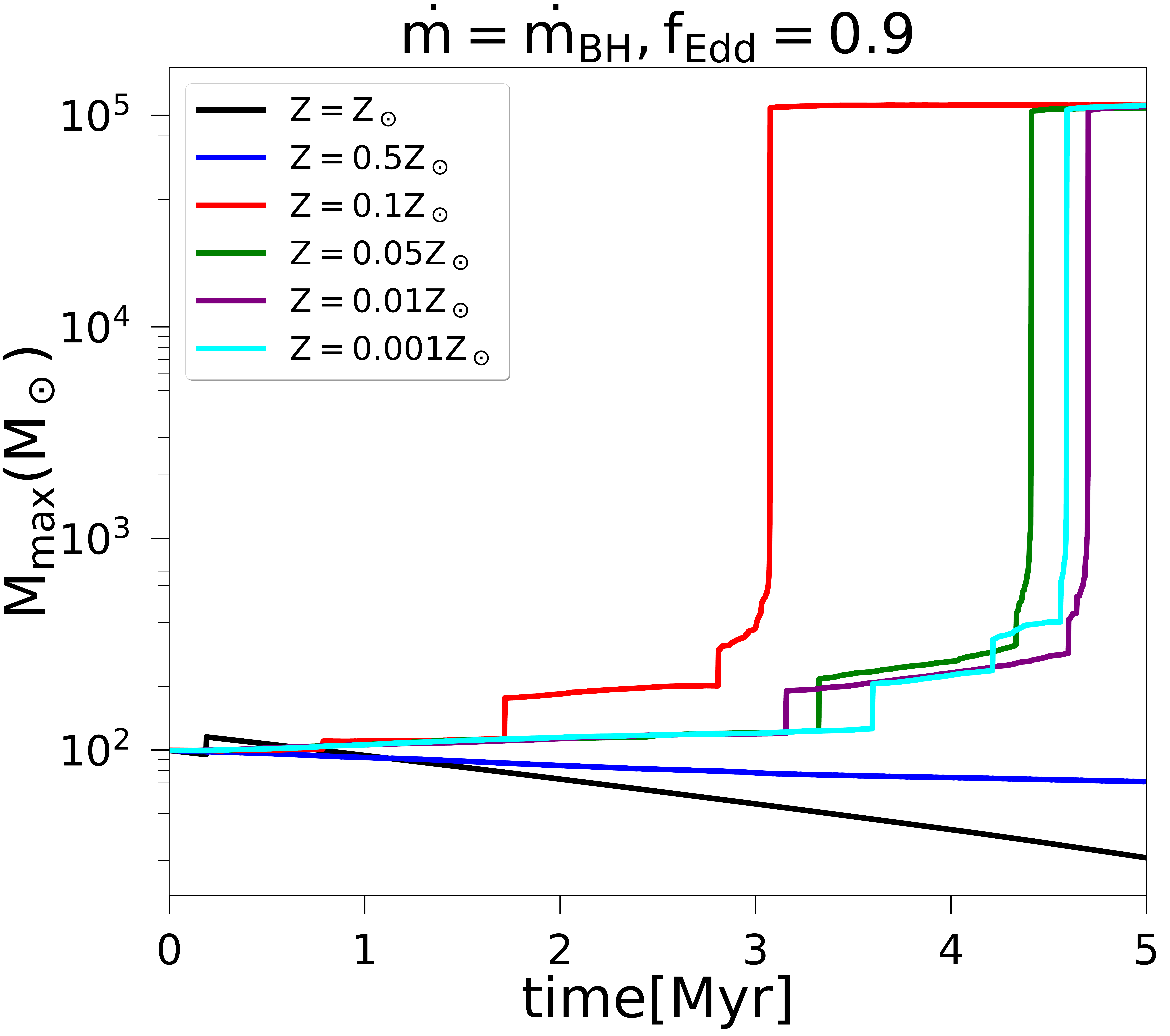}
    \caption{Mass evolution of the MMO for different accretion rates and mass loss rates. Different colors represent different values of $Z_\ast$ as labeled.}
    \label{metal}
\end{figure*}

The evolution of the mass of the MMO is shown in Fig.~\ref{metal} for constant accretion rates of $10^{-5}\Mdot$ and $10^{-4}\Mdot$, as well as for physically-motivated accretion rates of Eddington and Bondi-Hoyle accretion rates. For a constant accretion rate of $10^{-4}\Mdot$, the growth of the MMO is quite rapid due to mergers of collision products. The mass of the MMO reaches $\sim10^4 \MSun$ already after $0.8$ Myr except for $Z_\ast=\ZSun$ for $\fEdd=0.7$ and 0.9. The final mass of the MMO depends on $Z_\ast$ and $\fEdd$. SMSs of mass $\sim 10^4\MSun$ are formed for all the cases except for $Z_\ast=\ZSun$ for $\fEdd=0.7$ and 0.9 where the final mass of the MMO is $\sim 5\times 10^3\MSun$. For the case of a constant accretion rate of $10^{-5}\Mdot$, the growth of the  MMO is more gradual. The MMO reaches a mass of $\sim 10^4\MSun$ for $Z_\ast< 0.5\ZSun$ after $\sim 2.5$ Myr, and a final mass of $\sim 2\times 10^4\MSun$ after 5 Myr. For the case of $Z=0.5\ZSun$, the evolution of the MMO depends on the adopted $\fEdd$. For $Z_\ast= 0.5\ZSun$, the MMO reaches a mass of $\sim 8\times10^3\MSun, 6\times10^3\MSun$ and $2\times10^3\MSun$, after $\sim 2.5$ Myr for $\fEdd=0.5,0.7$ and 0.9, respectively. The final mass after 5 Myr varies between $~3\times 10^3-10^4 \MSun$. For the case of $Z=\ZSun$, no SMS could be formed for $\fEdd=0.9$. However, for $\fEdd=0.5$ and 0.7, a significant growth occurs after 3 Myr, and the MMO reaches a mass of $\sim 5\times 10^2\MSun$ and $4\times 10^3\MSun$ for $\fEdd=0.7$ and 0.5, respectively, at 5 Myr. The mass loss by winds is stronger for higher metallicity, and so its effect on the final mass of the MMO is more prominent in the case of the lower accretion rate $10^{-5}\Mdot$ compared to $10^{-4}\Mdot$.

Next, we explore the Eddington accretion scenario given by Eq.~\ref{eddacc}. For the case of Eddington accretion, the growth of the MMO occurs after about 3.5 Myr. However, it is important to note that using this recipe, the MMO does not grow for the cases of $Z_\ast=\ZSun$ and $0.5\ZSun$. The final mass of the MMO depends highly on the values of $Z_\ast$ and $\fEdd$. For $\fEdd=0.5$, an MMO of mass $\sim 10^3\MSun$ is formed for $Z_\ast \lesssim 0.01\ZSun$. For $\fEdd=0.7$ and 0.9, an MMO of mass $\sim 10^3\MSun$ is formed for $Z_\ast \lesssim 0.001\ZSun$. Finally, we explore the more extreme case of Bondi-Hoyle accretion given by Eq.~\ref{mainbondi}. The growth of the MMO is very slow during the initial period of time (depending on $Z_\ast$ and $\fEdd$), after which the growth happens in a runaway fashion due to the fact that $\dot{M}_\mathrm{BH}\propto M_\ast^2$. The timing of the runaway growth depends on when the first collision happens. Similar to the Eddington accretion scenario, there is no growth of the MMO for the case of $Z=\ZSun$ and $0.5\ZSun$. The MMO reaches a final mass of $\sim 10^5\MSun$ for $Z_\ast=0.1\ZSun$ or lower. The results could also be understood from the comparison of accretion and mass loss rates in Fig.~\ref{rates}. For a constant accretion rate of $10^{-4}\Mdot$, the accretion rate is greater than the mass loss rate for $M_\ast\lesssim 10^3\MSun$. So the stars have a net gain of mass no matter what metallicity we choose and as a result, they slow down and move towards the core due to momentum conservation. This results in a significant number of collisions and the formation of a SMS in the core. However, for the accretion scenario of $10^{-5}\Mdot$, Eddington, or Bondi-Hoyle, the accretion rate could be greater or lesser than the mass loss rate depending on the adopted values of $Z_\ast$ and $\fEdd$.

\section{Neglected processes}\label{neglected}
In this paper, we considered the interplay of collisions, physically motivated accretion recipes, and mass loss due to stellar winds. However, there are important processes that were still neglected, and which could have a relevant influence on some of the results.

In the context of stellar winds, we considered only the mass loss, but the winds also deposit kinetic energy into the system. It is therefore important to at least approximately assess its effect.
\begin{figure}
    \includegraphics[width=\columnwidth]{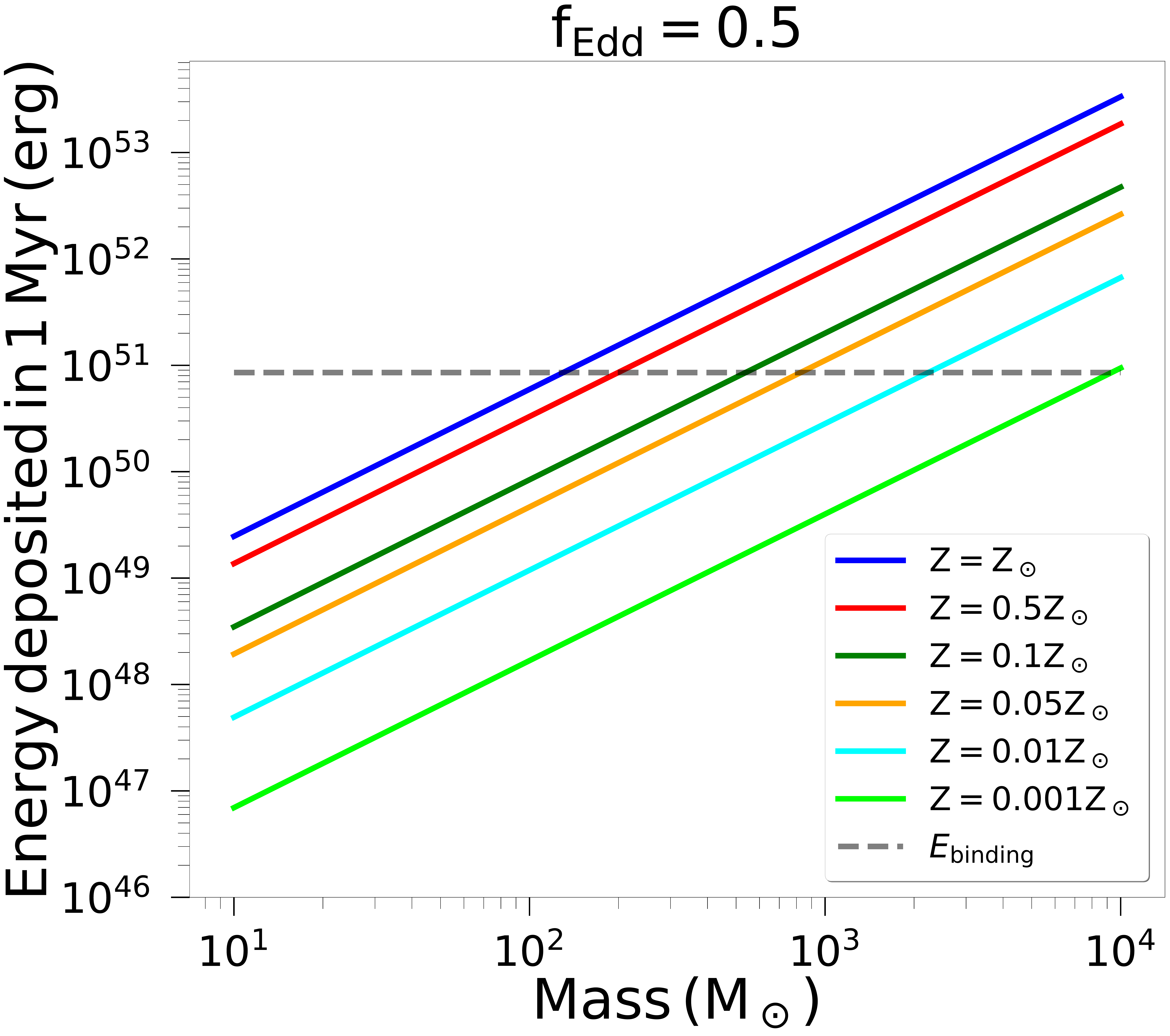}
    \caption{Energy deposited by a single star as a function of mass. Different color represents different $Z_\ast$. The binding energy of the cluster is shown as the black dashed line for comparison }
    \label{deposit}
\end{figure}
In Fig.~\ref{deposit}, we show the energy deposited by a single star for different metallicities as a function of mass. The energy deposited by a single star can be calculated as $\dot{E}_{\rm \ast, kin}\sim\dot{M}_{\rm loss}v^2$, where $\dot{M}_{\rm loss}$ is the mass loss rate of the star (computed for $\fEdd=0.5$). To estimate the velocity of the winds we calculate the escape velocity from the stellar surface, $v_{\rm esc}=\sqrt{2GM_\ast/R_\ast}$. The wind velocity should correspond to that velocity within a factor of a few \citep{Vin00,Vin01}. We also show the gravitational binding energy of the cluster  $E_\mathrm{bin} \simeq GM^2/R\sim9\times10^{50}$~erg with the black dashed line in the same figure for comparison. It is important to note that a single star with mass $\sim$ few times 100 $\MSun$ will deposit enough energy in 1 Myr to unbind the cluster for metallicities $Z_\ast\gtrsim 0.5\MSun$. As we go toward lower metallicities, the energy deposition rate is considerably lower. In order to avoid the unbinding of the cluster, one can naively expect the stars to be of lower metallicities. For a constant accretion rate of $10^{-4}\Mdot$ the SMS is forming within the first 1 Myr, so in principle the deposition of energy by the stellar wind will not prohibit the formation of a SMS. However, for an accretion rate of $10^{-5}\Mdot$ the SMS is forming at a much later stage, $\gtrsim 2$ Myr. The energy deposited by the wind from the MMO is enough to unbind the cluster within $~2$ Myr for $Z_\ast\gtrsim 0.1\ZSun$. For the Eddington accretion scenario, the SMS $\sim 10^3\MSun$ is forming at a much later stage $\sim 4-5$ Myr for $Z\lesssim 0.01$ Myr for which the binding energy of the cluster is greater than the kinetic energy deposited by the star. So, the deposition of kinetic energy from the star would not be expected to be a problem at least for the run-time of the simulations. For the Bondi-Hoyle scenario, no SMS is forming for $Z_\ast\gtrsim 0.5\ZSun$, where we expect the unbinding of the cluster to be much faster. For all other cases, the SMS could be formed if kinetic energy feedback was not relevant; however in reality it will unbind the cluster very quickly ($\lesssim 1$ Myr), since for Bondi-Hoyle the mass of the SMS is $\sim 10^5\MSun$.
 
To compute the energy deposited by the whole cluster we can assume a simplified cluster with $N=5000$ stars with each star of mass $22\MSun$ (which is the average mass of a star in the cluster with the initial conditions we assumed in our simulations). The velocity of the wind can be estimated as for a single star, which yields a characteristic velocity $v\sim 1100$~km/s.  The total kinetic energy deposition rate can then be evaluated as $\dot{E}_{\rm kin}\sim\dot{M}Nv^2\sim 1.2 \times10^{53}$~erg Myr$^{-1}$.

To avoid unbinding the gas within a timescale of $1$~Myr, the energy deposition should decrease by at least two orders of magnitude. Equation~(21) of \citet{Vin01} suggests that the mass loss rate scales with the metallicity as $Z^{0.85}$, implying that a decrease of the metallicity by a factor of $225$ should bring the energy deposition rate into the regime where the gas no longer becomes unbound. Since this estimate is very approximate, we expect the transition where kinetic energy deposition is no longer relevant to occur somewhere in the range ($10^{-2}-10^{-3})\,\ZSun$. For the range in between, gas expulsion due to the winds is expected to limit the potential growth of the central massive object, with this effect becoming weaker at lower metallicities. We also want to note that the gravitational potential energy of the cluster will change with the cluster properties. For a massive cluster or for a compact cluster the binding energy will be higher and hence the formation of the SMS would be more favorable. 

On the $5$~Myr timescale explored in our simulations, supernova feedback is also expected to become relevant. With the typical energy of $10^{51}$~erg for a core collapse supernova, it is clear that one such event will expel the gas and terminate the accretion, if it has not stopped already (either due to gas expulsion by stellar winds or as the accretion process may have depleted the gas).

In future work, it will be important to study detailed gas dynamics where the kinetic energy deposition of winds as well as the supernova feedback is taken into account.

\section{Summary and Discussion}\label{Discussion}
In this work, we explored the effect of mass loss due to stellar winds on the final mass of the SMSs, which could be formed via runaway stellar collisions and gas accretion inside NSCs. We find that a SMS of mass $\gtrsim 10^3\MSun$ could be formed even in a high metallicity environment for high accretion rates of $10^{-4}\Mdot$. For an accretion rate of $10^{-5}\Mdot$, the final mass of the SMS $\sim 10^4\MSun$ for $Z_\ast\lesssim 0.5\ZSun$. Whereas for solar metallicity, no SMS can be formed for $\fEdd=0.9$ and SMSs of masses $\sim 10^{2-3}\MSun$ can be formed for $\fEdd=0.7$ and 0.5, respectively. For the case of Eddington accretion it will not be possible to form a SMS in the metallicity regime $\gtrsim 0.1\ZSun$. Finally, for the Bondi-Hoyle accretion scenario, we find that the formation of a SMS will not be possible in the high metallicity regime of $Z_\ast\gtrsim 0.5\ZSun$. 

The interaction of the stellar wind and the gas inside the cluster might play an important role in the evolution of the SMS. The winds from the SMSs have high velocities $\sim 10^3$ km\,$\mathrm{s^{-1}}$~\citep{Mui12}, which might exceed the escape velocity from the centre of our modelled star cluster. The SMS in the cluster is close to the centre which results in a high collision rate near the centre due to a shorter relaxation time in the core and an increased collisional cross section. If the SMS is displaced by collisions, it rapidly sinks back close to the centre via dynamical friction where it may eventually decouple from the remainder of the cluster. This is also known as the Spitzer instability~\citep{Spi69}. Interestingly,~\citet{Krau16} have found that for a Salpeter type mass function the stellar wind cannot remove the gas inside the cluster. Hence, we do not expect the stellar wind to remove gas from the cluster.

One of the main caveats of this work is the neglect of the kinetic energy deposition through stellar winds, which could contribute significantly to expel the gas. The latter is likely to create a regime where the growth of a massive object is still inhibited, even though the mass loss itself from the winds is no longer significant. Below a critical metallicity in the range ($10^{-2}-10^{-3})\,\ZSun$, this effect is no longer expected to be relevant; however, supernova feedback may  lead to the expulsion of the gas. Another relevant caveat is the extrapolation of the mass loss recipe by~\citet{Vin00,Vin01} beyond $1000\MSun$, for which the mass loss is not really well known. Another uncertainty is the mass loss rates for stars close to their Eddington limit. \citet{Vin11} have shown  that the mass-loss rate increases strongly for stars close to the Eddington limit. So we might be underestimating the mass loss rate assuming the \citet{Vin01} recipe, especially in the high mass regime. We point out that similar to~\citet{Das20}, this work contains an idealized simulation setup. In real cosmological systems, the gas dynamics could be different and one needs to solve the hydrodynamics equations. In order to study the gas dynamics in detail, we need to incorporate the full hydrodynamics and hence the cooling, which also depends on the chemistry of the gas. Feedback processes due to the stars would also need to be modeled in more detail. Using cosmological zoom-in simulations,~\citet{Li17} have found that accretion might be regulated by stellar feedback processes. The main goal of this work was to build a simplified model that allows us to study the evolution over a large part of the parameter space for a long timescale of a few Myr. For future work, it will be important to explore more realistic accretion scenarios and their interaction with the mass loss process, as well as mass loss in the range of high stellar masses.

\section*{Acknowledgements}
We thank the anonymous referee for constructive comments on the manuscript. This work received funding from the Mitacs Globalink Research Award, the Western University Science International Engagement Fund, the Millenium Nucleus NCN19$\_$058 (TITANs) and BASAL Centro de Excelencia en Astrofisica y Tecnologias Afines (CATA) grant PFB-06/2007. This research was made possible by the computational facilities provided by the Shared Hierarchical Academic Research Computing Network (SHARCNET: www.sharcnet.ca) and Compute Canada (www.computecanada.ca).
This project was supported by funds from the European Research Council (ERC) under the European Union’s Horizon 2020 research and innovation program under grant agreement No 638435 (GalNUC), and a Discovery Grant from the Natural Sciences and Engineering Research Council (NSERC) of Canada.

\section*{Data availability}
The data underlying this article will be shared on reasonable request to the corresponding author.

\bibliographystyle{mnras}
\bibliography{nuclearcluster} 





\bsp	
\label{lastpage}
\end{document}